\documentclass{customized}

\usepackage{graphicx}
\usepackage[percent]{overpic}
\usepackage{epstopdf, epsfig}
\usepackage{graphicx}
\usepackage{dcolumn}
\usepackage{bm}
\usepackage{color}
\usepackage{array}
\usepackage{amsmath,amssymb,stmaryrd} 
\usepackage{multirow}
\usepackage{booktabs,comment}
\DeclareMathOperator\erf{erf}

\definecolor{blue3}{rgb}{0, 0.1770, 0.3410}

\begin{document}  

\newtheorem{lemma}{Lemma}
\newtheorem{corollary}{Corollary}

\shorttitle{$Re$ effects on wake of wavy cylinder} 
\shortauthor{K. Zhang \textit{et al.}} 

\title{Reynolds number effects on the bistable flows over a wavy circular cylinder}

\author
 {
 Kai Zhang\aff{1,2}\corresp{kai.zhang@sjtu.edu.cn}
, Hongbo Zhu\aff{1}, Yong Cao\aff{1} \and Dai Zhou\aff{1}\corresp{\email{zhoudai@sjtu.edu.cn}}}

\affiliation
{
\aff{1}State Key Laboratory of Ocean Engineering, School of Naval Architecture, Ocean and Civil Engineering, Shanghai Jiao Tong University, Shanghai 200240, China
\aff{2}Department of Mechanical and Aerospace Engineering, Rutgers University, Piscataway, NJ 08854, USA
}

\maketitle

\begin{abstract}
The wake of wavy cylinder has been shown to exhibit bistability.
Depending on the initial condition, the final state of the wake can either develop into a steady flow (state I), or periodic shedding (state II).
In this paper, we perform direct numerical simulations to reveal the Reynolds number effects on these two wake states.
With increasing Reynolds number, the steady vortical structures in state I wake sways back and forth in the spanwise direction, resulting in low-frequency fluctuations in drag forces, but not in lift.
For state II, the increase in Reynolds number is associated with the emergence of another spectral peak in the lift coefficient.
The secondary frequency is associated with highly three-dimensional vortical structures in the wake.
For both states, the wakes transition to turblent flows at higher Reynolds numbers, with the development of small-scale vortices.
We further study the streamwise gust flows over the wavy cylinder.
The time-varying inflow velocity results in a wide range of instantaneous Reynolds number spanning from the absolutely unstable flow regime to the bistable regime.
Depending on the period of the inflow velocity variation, the wake perturbations grown at the absolutely unstable flow regime can be damped out in state I wake, or grow large enough to trigger the transition state II, resulting in loss of flow control efficacy.
The above analyses reveal novel flow physics of the bistable states at unexplored Reynolds numbers, and showcase the complex transition behavior between the two states in unsteady flows. 
The insights gained from this study improve the understanding of the wake dynamics of the wavy cylinder.

\end{abstract}

\section{Introduction}

Engineering structures featuring bluff bodies, such as deepwater riser, long-span bridge cables are often plagued by the excessive fluid-induced forces and vibrations, which increase the construction cost, and cause concerns of fatigue failure.
In order to alleviate these detrimental effects, a lot of efforts have been devoted to the manipulation of flow behind bluff structures \citep{choi2008control,kumar2008passive,triantafyllou2016vortex}.
In particular, spanwise modification near the separation point, referred as 3D forcing \citep{choi2008control}, has been proven effective for the reduction of hydrodynamic forces on nominally two-dimensional bluff bodies.
Typical examples of this category include sinus axis and helical bumps \citep{owen2001passive}, small-size tabs \citep{yoon2005control}, spanwise blowing and suctions \citep{kim2005distributed}, just to name a few.

Among the various realizations of the 3D forcing techniques, a circular cylinder with straight axis and axially varying diameter (referred to as wavy cylinder hereafter), has attracted a lot of attention due to its omnidirectional geometry.
Early experimental investigations by \citet{ahmed1992transverse} and \citet{ahmed1993experimental} have shown that pressure gradient exists in the spanwise direction, leading to the formation of non-uniform separation lines and the development of the three-dimensional wake. 
The subsequent investigations discovered that such a three-dimensional wake is associated with significant suppression of the K\'arm\'an vortex shedding, and reduction in drag and lift forces\citep{lam2009effects,xu2010large,lam2008large,lin2016effects,jung2014large}.
Moreover, the spanwise-undulated geometry is found to resemble the whiskers of harbor seals \citep{hanke2010harbor,lyons2020flow,murphy2021other}. 
Such shape has been shown to exhibit superior hydrodynamic performance and to enhance the sensitivity of the whisker even in turbid water \citep{beem2015wake}. 
This finding has inspired bio-mimicry innovations such as energy conserving flow sensors \citep{beem2012calibration}.

Despite the favorable aerodynamic properties regarding the wavy cylinder as described above, we have revealed in a series of studies that the flow control efficacy of the wavy cylinder, even optimally designed to eliminate the K\'arm\'an vortex shedding, can be lost in some occasions.
For example, when the wavy cylinder is yawed to the incoming flow, the K\'arm\'an vortex shedding that is concealed in the non-yawed condition revives, leading to increased drag and fluctuating lift \citep{zhang2018large}. 
When flexibly mounted, our numerical simulations found that the same wavy cylinder can also develop large-amplitude vortex-induced vibrations similar to a straight cylinder \citep{zhang2017numerical}.
This observation is supported by the experimental study by \citet{assi2018vortex}.
In addition, if the wavy cylinder is forced to oscillate in the crossflow direction, the K\'arm\'an vortex shedding mode can revive, leading to quasi-periodicity in the flow, as well as synchronization between the inherent shedding frequency and the forcing frequency \citep{zhang2018numerical}.

\begin{figure}
\centering
\includegraphics[width=0.6\textwidth]{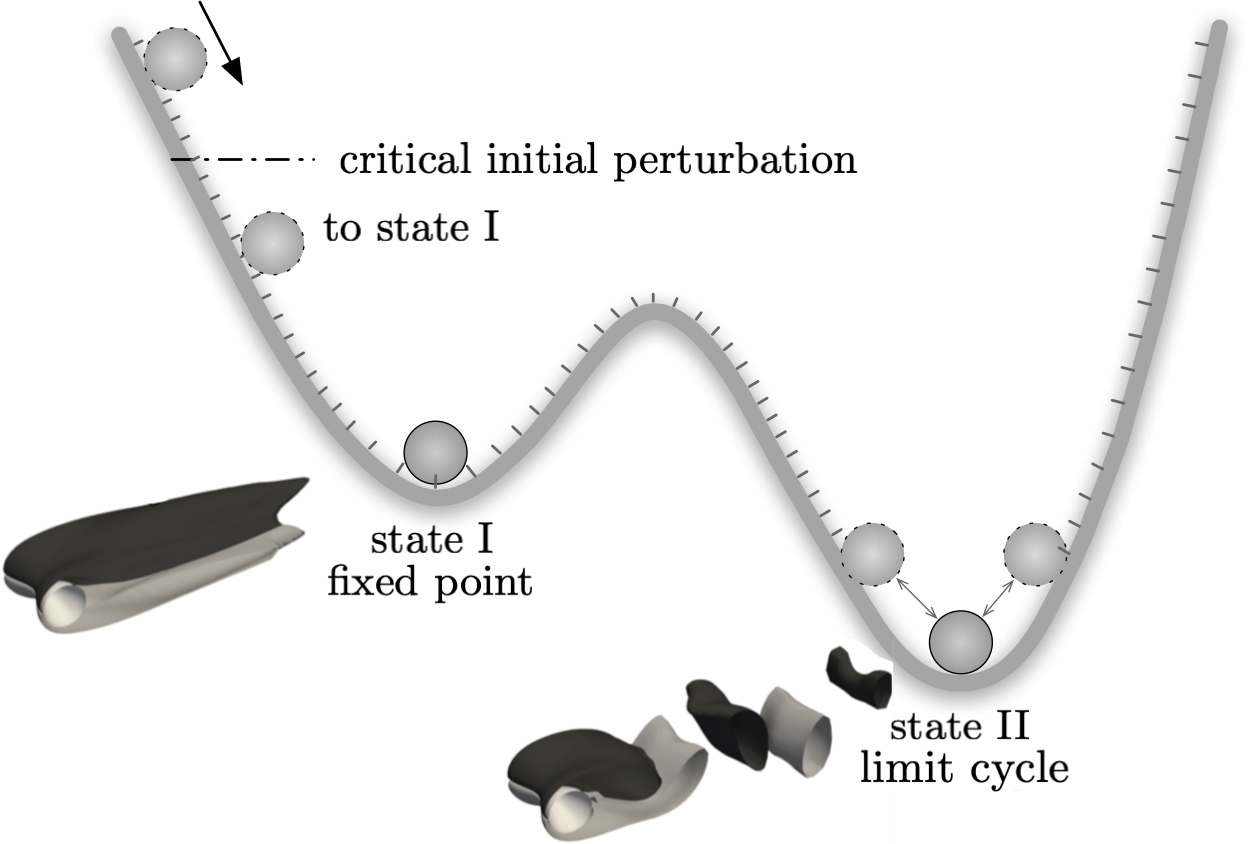}
\caption{Schematic diagram of the bi-stable states of the wavy cylinder wake at $Re=150$. State I represents the steady wake, and state II represents the oscillatory wake. Reproduced from \citet{zhang2020bistable} with permission obtained from AIP Publishing.}
\label{fig:bistableScheme}
\end{figure}

The above observations hint that K\'arm\'an vortex shedding mode is inherent to the wavy cylinder wake.
Indeed, in our latest work \citep{zhang2020bistable}, the bistability in the wake of the wavy cylinder was unveiled.
It was shown that, even in the fixed configuration, the final state of the wavy cylinder wake depends on the initial condition.
As depicted in figure \ref{fig:bistableScheme}, for weakly perturbed initial condition (such as uniform initial condition), the wake settles to the fixed point featuring steady flow with low drag and zero lift (state I).
On the other hand, if the initial condition is highly perturbed, the wake finally arrives at the limit cycle state featuring K\'arm\'an vortex shedding (state II), indicating the loss of flow control efficacy.
However, our previous work was limited to low Reynolds number of $Re\approx 150$.
The wake dynamics of the wavy cylinder at higher Reynolds numbers is not systematically explored.

In this paper, we extend our previous works to study the Reynolds number effects ($Re$ = 30 -- 300) on the bistable flows over the wavy cylinder.
The aim is to discover new flow phenomena in both wake states in the unexplored $Re$ regimes.
In addition, we simulate time-varying flows over the wavy cylinder to reveal the complex transition behaviors between the two wake states.
The insights gained from this study should improve the current understanding of the wake dynamics of the wavy cylinder, and potentially aid its use in engineering applications.
In what follows, we present the computational setup in \S \ref{sec:setup}.
The results are discussed in \S \ref{sec:results}, where we unveil some unique flow phenomena with increasing Reynolds number, and shed light upon the transition between the two states in streamwise gust flows.
We conclude this paper by summarizing the findings in \S \ref{sec:conclusions}.

\section{Computational setup}
\label{sec:setup}
\begin{figure}
	\centering
	\includegraphics[scale=0.5]{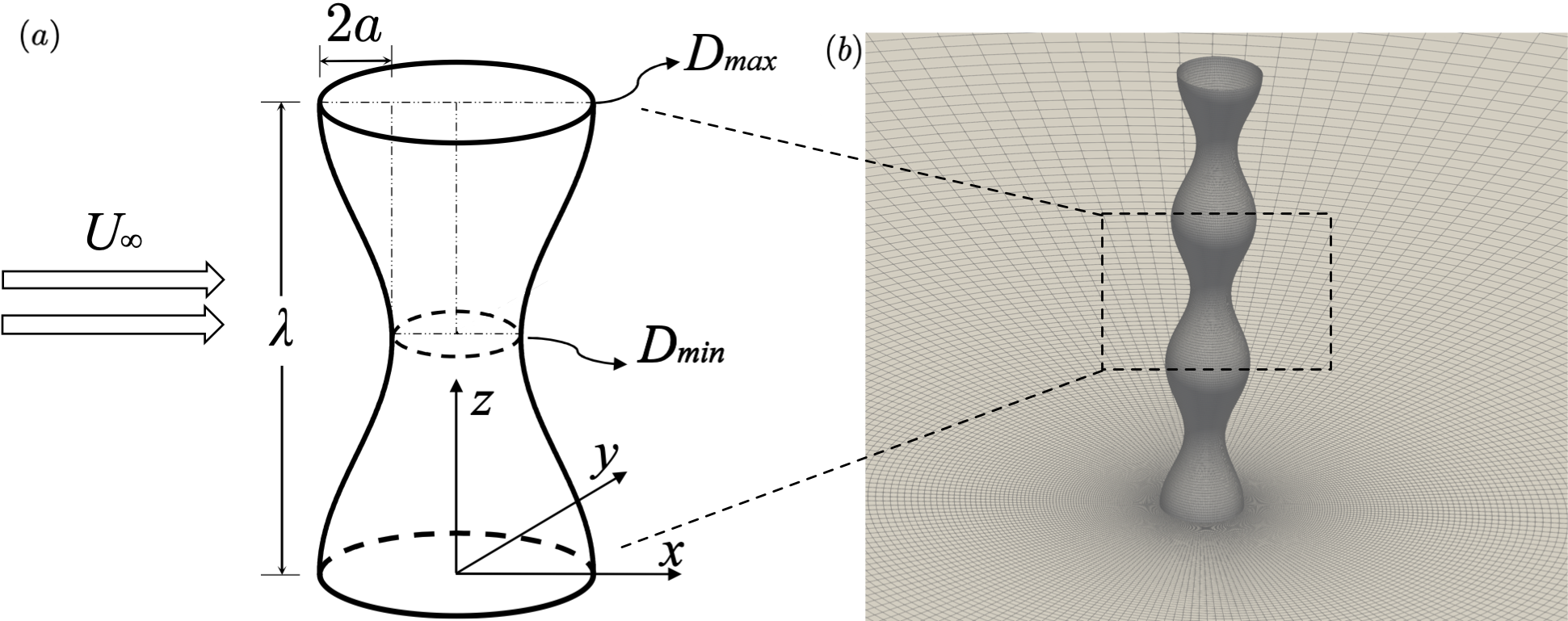}
	\caption{($a$) Schematic view of problem setup and $(b)$ mesh design.}
	\label{fig:scheme}
\end{figure}

The geometry of the wavy cylinder is schematically depicted in figure \ref{fig:scheme}. 
The diameter of the wavy cylinder varies sinusoidally along the spanwise direction $z$ according to
\begin{equation}
D(z)=D_m+a\cos(2\pi z/\lambda),
\end{equation}
where $D_m$ is the averaged diameter, $a$ and $\lambda$ are the geometric amplitude and wavelength, respectively. In the current paper, we assign $a=0.175D_m$ and $\lambda=2.5D_m$. This set of parameters has been shown by \citet{lam2009effects} and \citet{zhang2020bistable} to exhibit satisfactory flow control efficacy.
The wavy cylinder is subjected to a uniform incoming flow $U_{\infty}$ in the $x$ direction. For non-dimensionalization, we normalize the spatial variables by the averaged diameter $D_m$, velocity by $U_{\infty}$, time by $D_m/U_{\infty}$, and frequency by $U_{\infty}/D_m$. The Reynolds number, defined as $Re\equiv U_{\infty}D_m/\nu$, where $\nu$ is the kinematic viscosity of the fluid, is varied from 30 to 300.

The flow is governed by the incompressible Navier-Stokes equations, which are solved by the direct numerical simulations using the open-source software OpenFOAM. 
Both the time and space are discretized with second-order accurate schemes. 
The wavy cylinder is placed in the center of a circular computational domain of $30D_m$ in radius.
Periodic boundary condition is specified at the spanwise ends of the domain. 
It is noted that at higher Reynolds numbers, the flows at the two spanwise boundaries may not be periodic.
Thus, we use three wavelengths ($7.5D_m$ in spanwise direction) in the spanwise direction to alleviate the unphysical boundary condition effects on the wake dynamics.
The cylinder surface is treated as no-slip wall. 
A uniform flow condition with freestream velocity $U_{\infty}$ is specified at the inlet.
In order to achieve the state I wake, we initialize the flow with uniform velocity.
For state II wake, the flow is initialized with the saturated unsteady flow at $Re=100$, at which the wake is absolutely unstable.
The flow statistics are collected after the unphysical transient initial flows are flushed out from the computational domain.

An O-type mesh with resolution $N_c\times N_r\times N_z=200\times 180 \times 180$ (where $N_c$, $N_r$ and $N_z$ represent the grids in the circumferential, radial and spanwise directions) is used for the domain discretization.
The mesh is concentrated in the vicinity of the cylinder to better resolve the near wake. The nondimensional time-step is set to be $\Delta t=0.01$.
Mesh dependency test is omitted here since it has been carefully done in our previous study \citep{zhang2020bistable}.

\section{Results}
\label{sec:results}
\subsection{Reynolds number effects on bistable flows}

\begin{figure}
\centering
\includegraphics[width=0.6\textwidth]{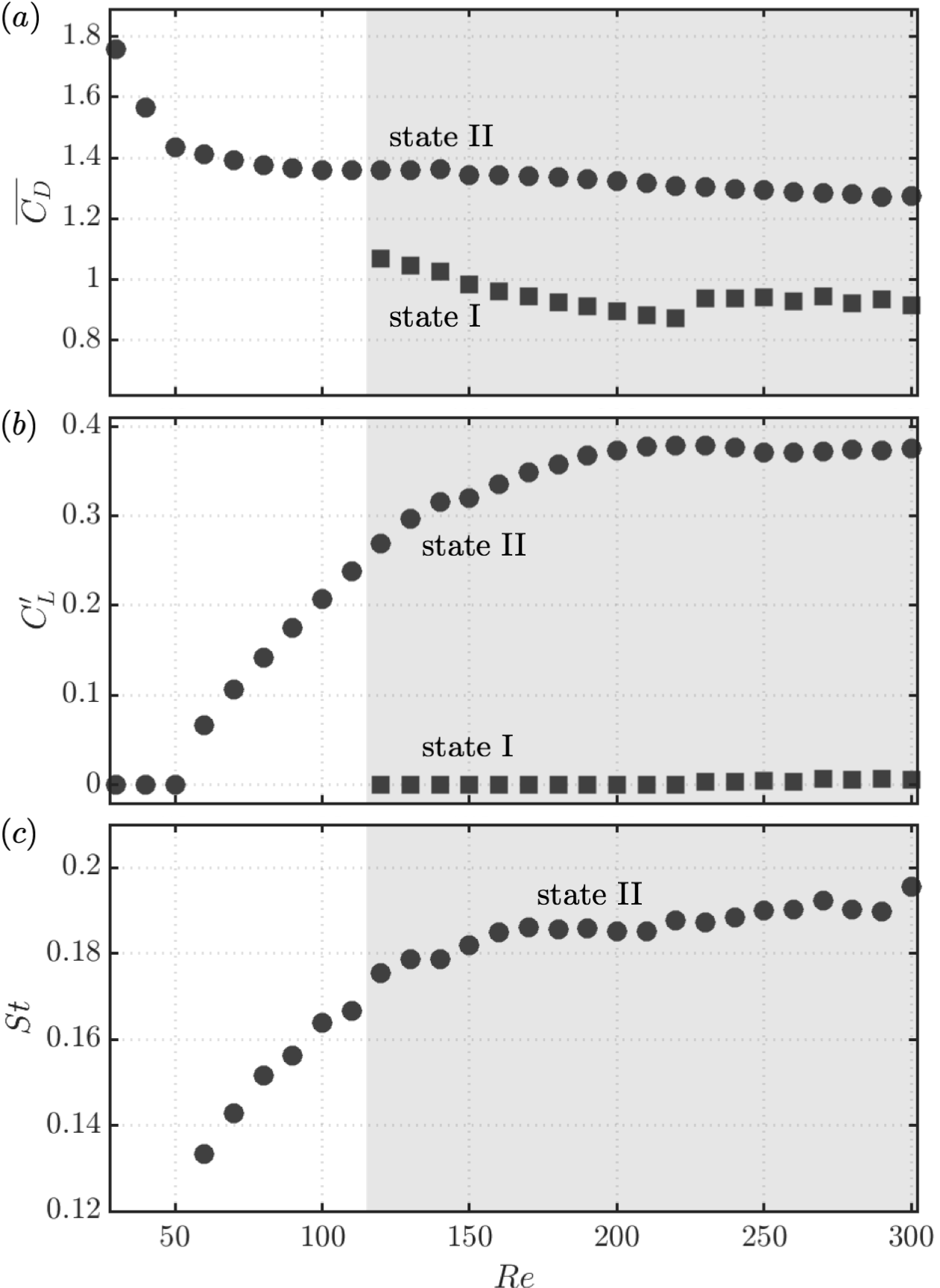}
\caption{($a$) Time-averaged drag coefficient, ($b$) rms lift coefficient, and ($c$) shedding frequency of the wavy cylinder. The shaded area indicates the bistable regime.}
\label{fig:CdCl}
\end{figure}

We first present an overview of the bistable flows from the perspective of aerodynamic forces of the wavy cylinder.
The time-averaged drag coefficients ($\overline{C_D}$), root-mean-squared (rms) lift coefficients ($C_L^{\prime}$), and the dominant shedding frequency $St$ are shown in figure \ref{fig:CdCl}.
For $Re\lesssim 110$, the flow over wavy cylinder exhibits similar dynamic features with that of the two-dimensional cylinder \citep{zhang2020bistable}.
The wake remains steady for $Re$ below 60, and becomes absolutely unstable with increasing Reynolds number.
The bistability commences at $Re=120$, as evidenced by the bifurcations in the $\overline{C_D}$-$Re$ and $C_L^{\prime}$-$Re$ curves.
The drag and lift coefficients in state I are significantly smaller than that in state II.
No dominant frequency is detected in the lift coefficients for state I wakes, because the lift is either exactly zero for $Re=120$ -- 220 (note that this does not mean the wake is absolutely steady), or feature small random variations at higher $Re$.

\begin{figure}
\centering
\includegraphics[width=0.9\textwidth]{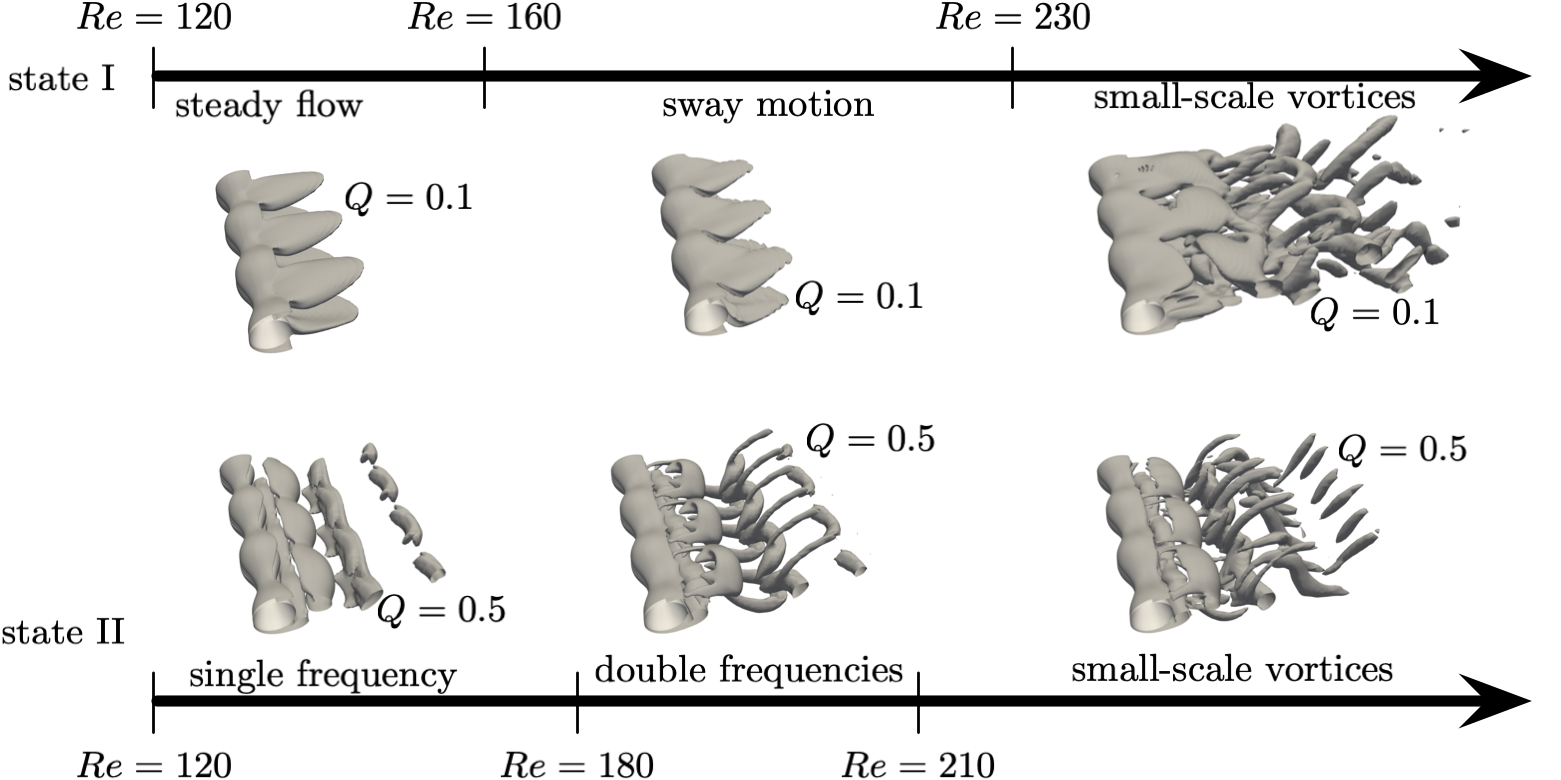}
\caption{Summary of Reynolds-number effects on the bistable flows over the wavy cylinder. The vortical structures are visualized by isosurfaces of $Q=0.1$ for state I, and $Q=0.5$ for state II.}
\label{fig:bistableRe}
\end{figure}

The representative vortical structures in the bistable regime are shown in figure \ref{fig:bistableRe}, with a preview of the Reynolds number effects.
For state I, the wakes at $Re=120-150$ are absolutely steady. 
The vortical structures visualized by $Q$ criteria feature a tongue-like structure that is located within each geometric wavelength.
In between $Re=160$ and 220, these tongue-like structures sways back and forth with low frequency, resulting in unsteadiness in the drag coefficients, but not in lift.
For $Re$ beyond 230, the vortex sheets breaks down in downstream, populating the far wake with small-scale vortices.
A more detailed analysis of the $Re$ effects on state I wake is presented in \S \ref{sec:stateI}.

For state II, the wake is primarily dominated by periodic vortex shedding.
The spanwise vortex tubes are more or less associated with waviness that is dictated by the geometric wavelength.
With increasing Reynolds number, the three dimensionality enhances in the wake.
This is also reflected in the lift spectra as increasing numbers of dominant frequencies, as will be discussed in \S \ref{sec:stateII}.

\subsubsection{Reynolds number effects on state I}
\label{sec:stateI}

\begin{figure}
\centering
\includegraphics[width=0.99\textwidth]{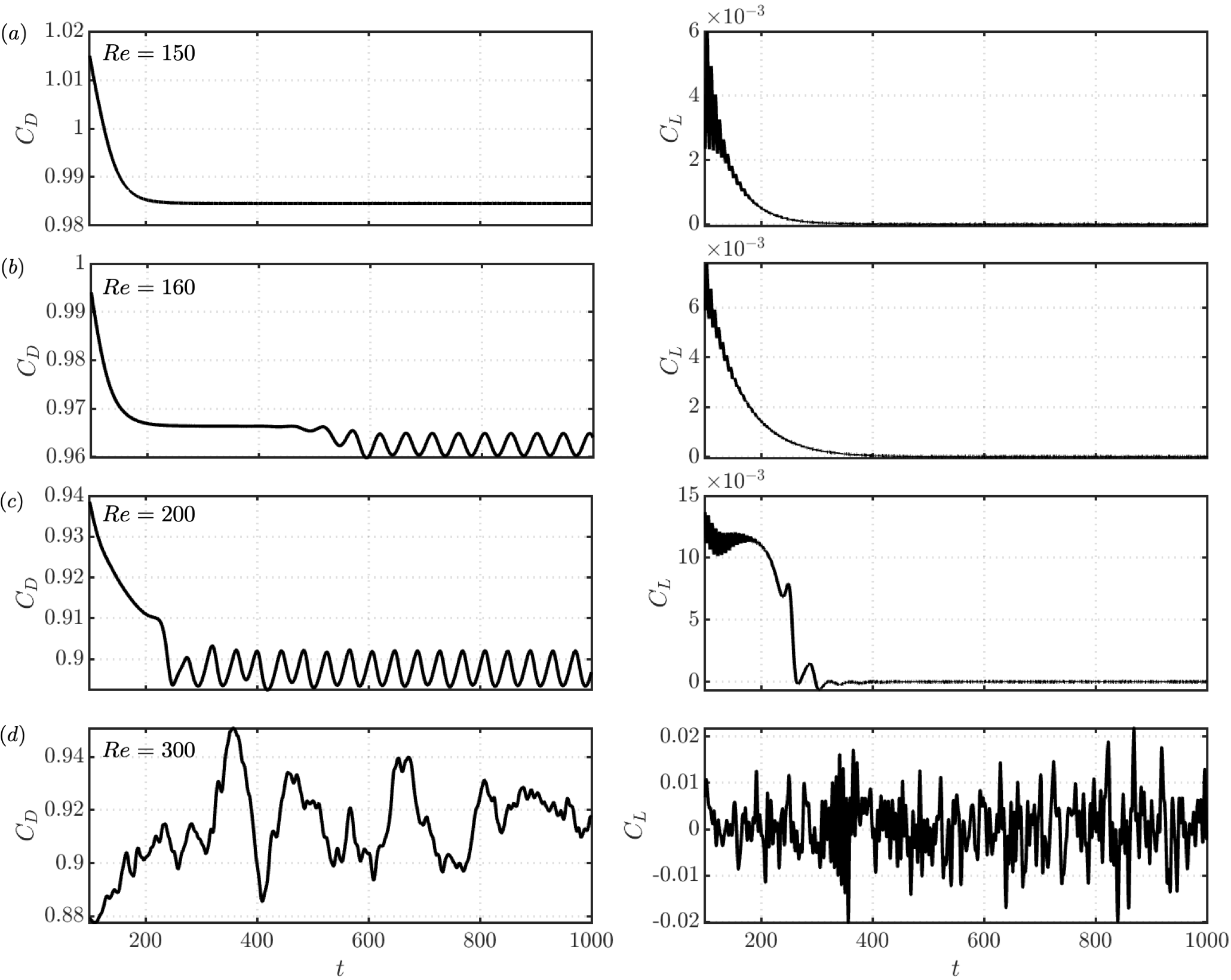}
\caption{Time histories of drag and lift coefficients of state I. $(a)$ $Re=150$, $(b)$ $Re=160$, $(c)$ $Re=200$ and $(d)$ $Re=300$.}
\label{fig:StateICdCl}
\end{figure}

In state I, the large-scale K\'arm\'an vortex shedding is suppressed in the wake.
However, with increasing Reynolds number, the wake can still exhibit unsteadiness.
We show the drag and lift coefficients of the wavy cylinder in state I wake at different Reynolds numbers in figure \ref{fig:StateICdCl}.
At $Re=150$, both drag and lift coefficients are constant after the initial transient are flushed out, suggesting the complete steadiness of the wake.
As $Re$ increases to 160, while the lift coefficient remains at zero, the drag coefficient exhibit periodic variation beyond $t\approx 500$.
It is noted that such phenomenon is associated with even smaller time-averaged drag than that in the steady flow regime.
Such phenomenon is also observed with increasing Reynolds number, although the time at which the unsteadiness in drag commences is much earlier than that at $Re=160$.
At higher Reynolds numbers, both drag and lift exhibit random unsteadiness, suggesting the emergence of small-scale vortices and the transition to turbulent flow.

\begin{figure}
\centering
\includegraphics[width=0.99\textwidth]{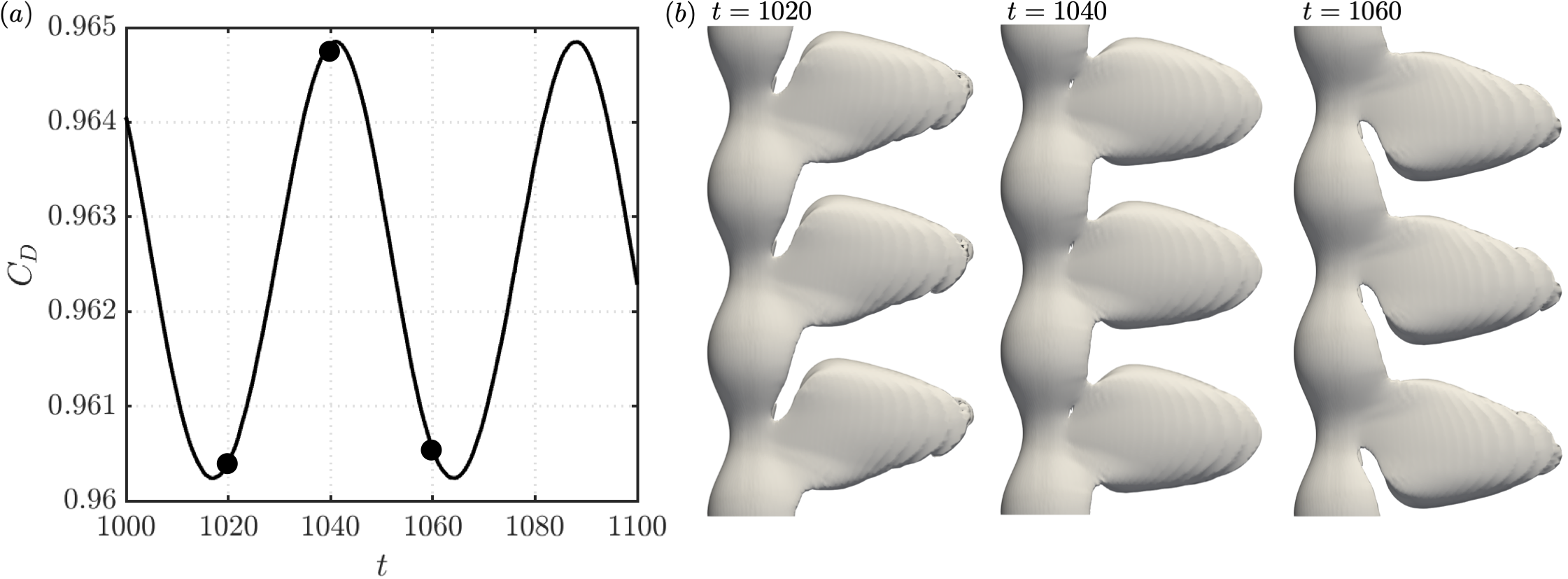}
\caption{Unsteady flow for the state I at $Re=160$. $(a)$ Drag coefficient and $(b)$ vortical structures visualized by isosurfaces of $Q=0.1$. The three time instants correspond to the black dots in ($a$). }
\label{fig:sway}
\end{figure}

To understand the unique unsteady flow at $Re=160$ -- 220 where the drag is unsteady but the lift is not, we plot the vortical structures of state I wake at $Re=160$ in figure \ref{fig:sway}.
As the drag coefficient varies periodically with time, the tongue-like vortical structure within each wavelength sways back and forth along the spanwise direction in sync.
Each time the vortical structures are positioned near the section of minimum diameter, the drag coefficient reaches maximum.
The minimum drag is reached when the vortical structures are positioned towards either sides. 
Thus, one cycle of vortical motion corresponds to two cycles in the drag variation.
Since the sway motions of the vortical structures are symmetric with respect to the $x$-$z$ plane, such unsteady flow does not induced variation in the lift coefficients.
Compared with the K\'arm\'an vortex shedding, the frequency of the vortical motion is much lower.
A summary of the period of the sway motion of the vortical structures is shown in table \ref{tab:swayPeriod}.
With increasing Reynolds number, the sway period decreases initially, and saturates at $T_{sway}\approx 80$ for higher $Re$.

\begin{table}
\caption{\label{tab:swayPeriod}Summary of the periods of the sway motion of the vortical structures}
\begin{tabular}{cccccccc}
\toprule
 $Re$	&	160	&	170	&	180	&	190	&	200	&	210	&	220 \\ \hline
 $T_{sway}$ & 94.1&  86.3 &   81.4&	80.5&	81.1&	81.7&	81.8\\  
\bottomrule
\end{tabular}
\end{table}

\subsubsection{Reynolds number effects on state II}
\label{sec:stateII}
\begin{figure}
\centering
\includegraphics[width=1\textwidth]{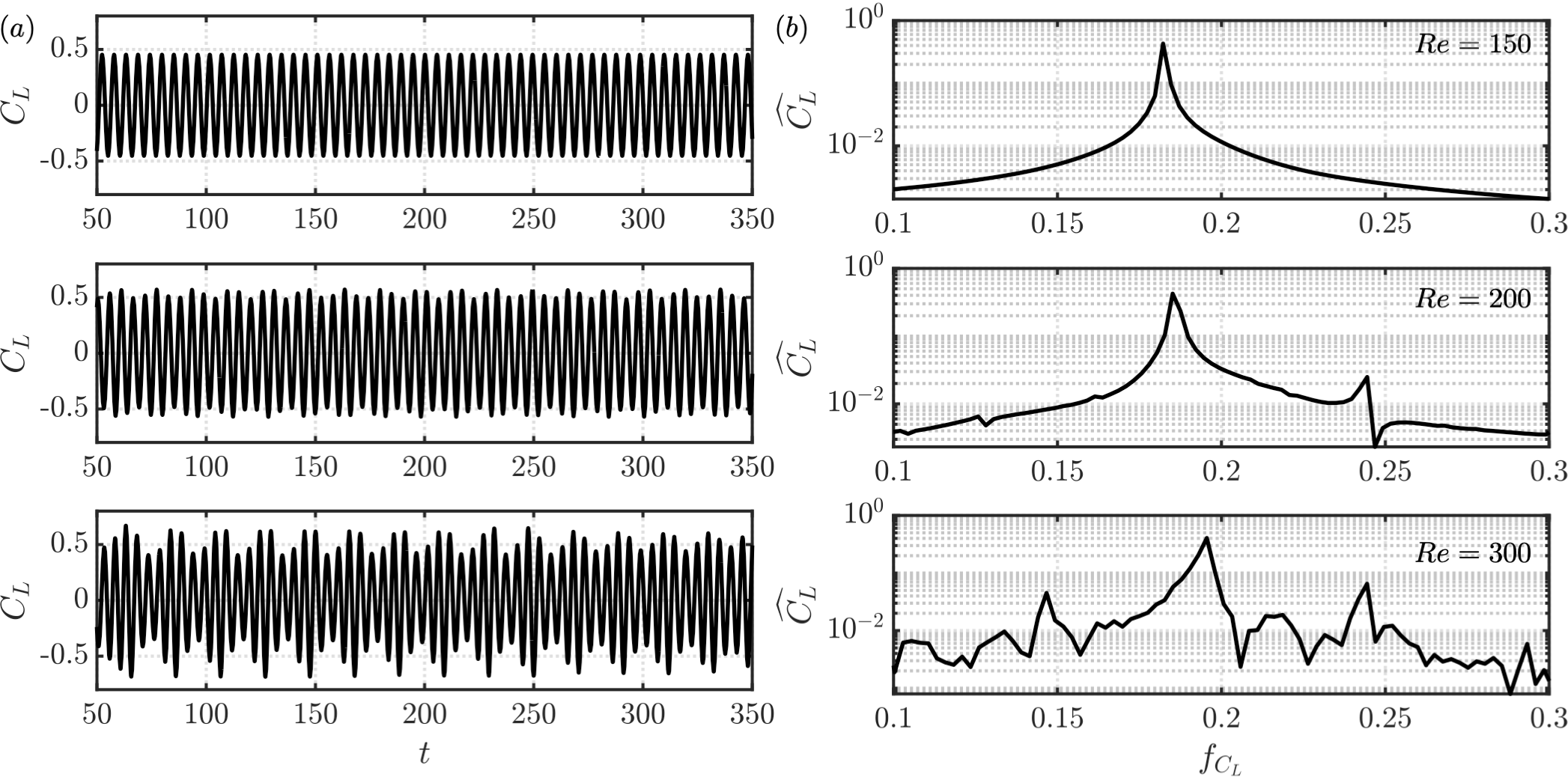}
\caption{($a$) Lift spectrum and ($b$) time history of lift coefficients for state II.}
\label{fig:stateIILift}
\end{figure}

The lift coefficients and their spectra of the state II wakes are shown in figure \ref{fig:stateIILift} for some representative Reynolds numbers.
At $Re=150$, the lift coefficient exhibit purely periodic oscillations dominated by a single frequency at $f=0.185$.
With increasing Reynolds number, the lift coefficient becomes quasi-periodic, and an additional frequency emerges at $f=0.244$, although its spectral amplitude is much weaker.
At higher Reynolds number, multiple peaks are observed in the lift spectrum, suggesting the flow is becoming turbulent.

\begin{figure}
\centering
\includegraphics[width=0.9\textwidth]{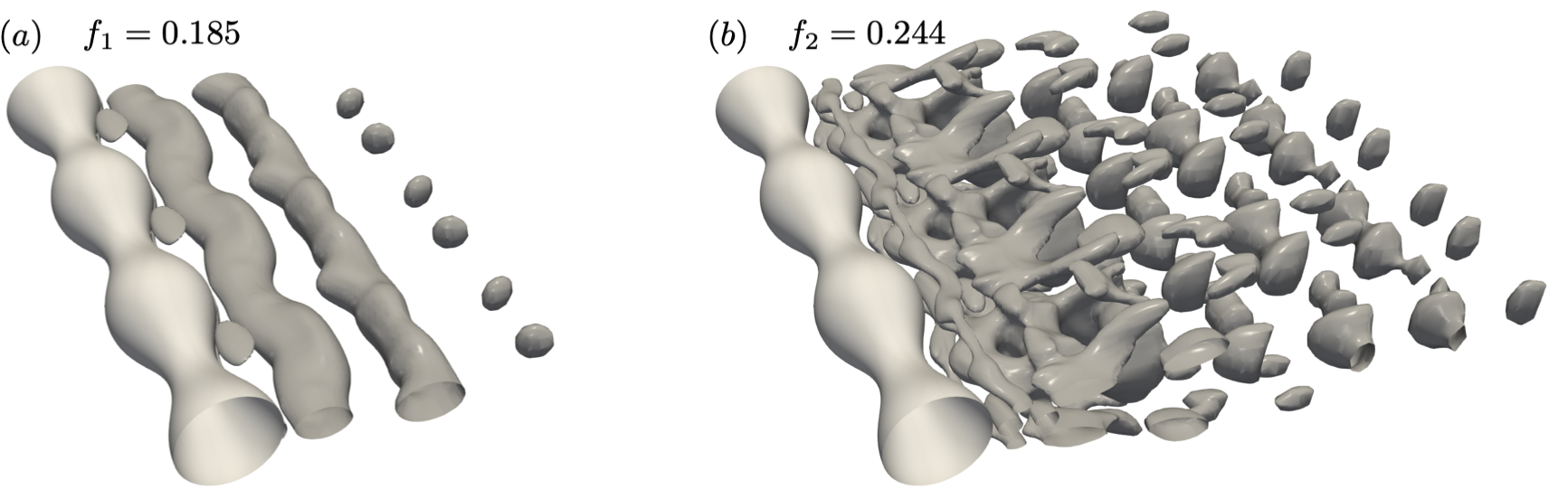}
\caption{Dynamic modes for the state II wake at $Re=200$. $(a)$  $f_1=0.185$ and ($b$) $f_2=0.244$. The two modes are visualized using the same contour level of $Q$ criterion.}
\label{fig:Re200_DMD}
\end{figure}

We use dynamic mode decomposition \citep{schmid2010dynamic,schmid2022dynamic} to reveal the coherent structures associated with the two prominent frequencies at $Re=200$.
As shown in figure \ref{fig:Re200_DMD}, for the primary shedding frequency $f_1=0.185$, the dynamic mode features spanwise coherent vortical structures with slight waviness.
Such structure resembles the typical two-dimensional K\'arm\'an vortex shedding mode, and is the dominant dynamic mode for the single-frequency unsteady flows at lower Reynolds numbers.
For $f_2=0.244$, the modal structure appears more complex as it involves both spanwise and streamwise vortical structures.
In addition, the vortical structures in this mode are also observed in the far wake, while the mode associated with $f_1$ is only strong in the near wake.
Nevertheless, the effects of the second mode on the lift of the wavy cylinder is much smaller than the previous one.

\subsection{Wake dynamics in time-varying inflow}
\label{sec:gust}
We further study the aerodynamics of the wavy cylinder in streamwise gust flows.
In order to save computational time, the spanwise length of the computational model is reduced to a single wavelength of $2.5D_m$ in these simulations.
A time-varying inflow velocity is specified at the inlet as
\begin{equation}
U_{in} = [\erf(0.02(t-300))+1] \cdot 0.35 \cdot \sin(2\pi t/T)+1.5,
\label{equ:gust}
\end{equation}
where $T$ denotes the period of the sinusoidal velocity variation.
With the Gaussian error function in equation (\ref{equ:gust}), the inlet velocity initially stay fixed, resulting in a constant Reynolds number of 150 up to $t=300$.
This allows the wake to develop into the steady state.
Beyond $t=300$, the velocity transitions to the sinusoidally varying part smoothly. 
The maximum instantaneous Reynolds number is 220, at which the flow exhibits bistability, and the minimum 80, at which the wake is absolutely unstable.

\begin{figure}
\centering
\includegraphics[width=0.99\textwidth]{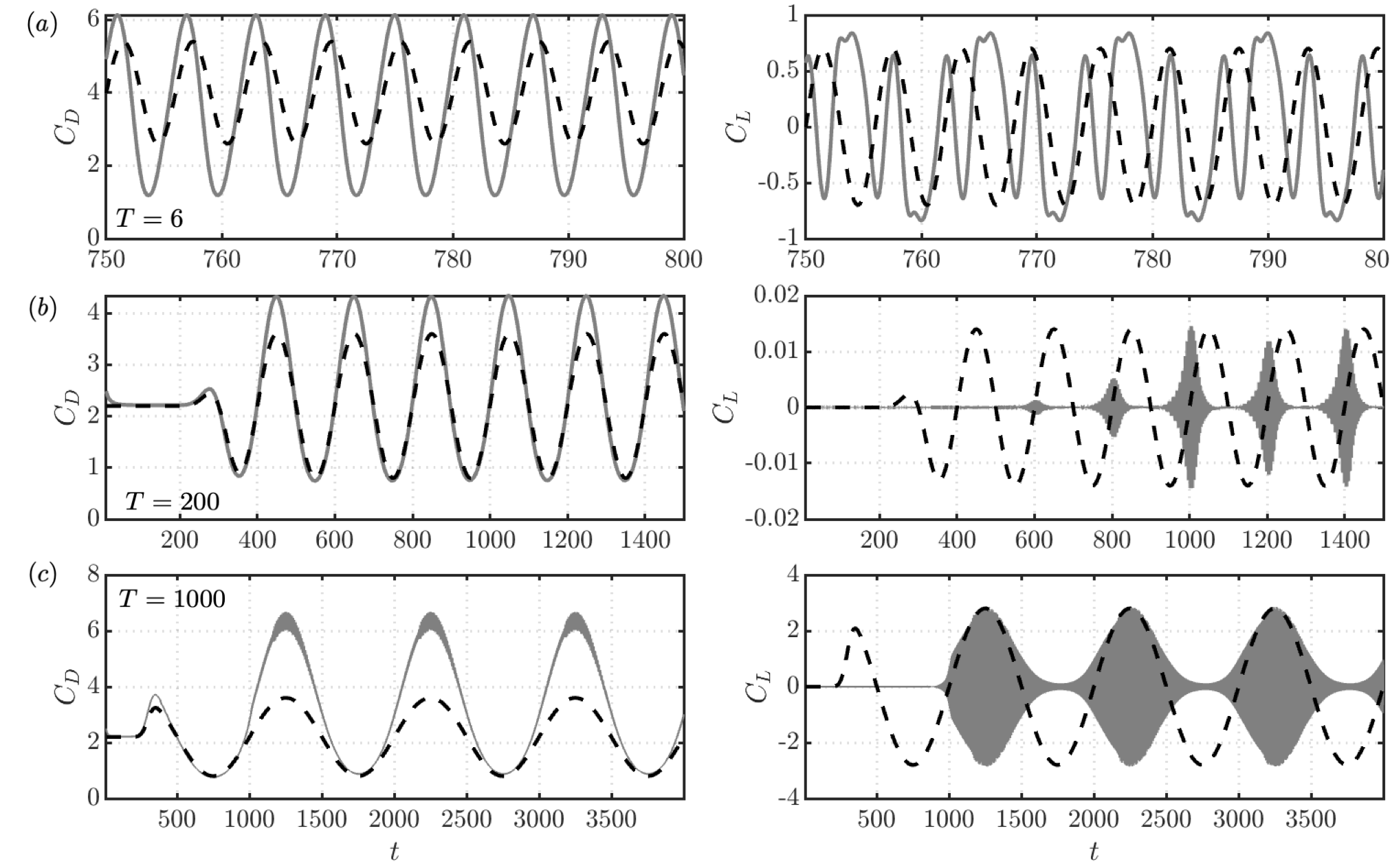}
\caption{Drag and lift coefficients of the wavy cylinder under time-varying inflow velocity. $(a)$ $T=6$, $(b)$ $T=200$ and $(c)$ $T=1000$. The gray solid lines are the force coefficients. The black dashed lines denote the variations of the inflow velocity, which results in a range of instantaneous Reynolds number from 80 to 220.}
\label{fig:gustCdCl}
\end{figure}

We assess the drag and lift coefficients of the wavy cylinder under different period $T$ in figure \ref{fig:gustCdCl}.
For $T=6$, which is of the same scale with the natural shedding period $T_{n}=5.55$ for the state II wake at $Re=150$, the dynamic inflow induces a quasi-periodic lift coefficient.
This is typical for flows over streamwise oscillating cylinders \citep{leontini2011numerical,kim2019lock}, which is a sister problem of the current varying velocity over a fixed cylinder.
This suggests that, despite the initially steady flow, the unsteady shedding state in the wavy cylinder's wake revives, and interacts with the velocity excitation, giving rise to the quasi-periodic flow behavior.
This observation is in accordance with our previous findings regarding the wake response of wavy cylinder under external forcings \citep{zhang2018numerical,zhang2020bistable}.

For $T=200$, which is much higher than natural shedding period, the lift coefficient remains low, and exhibits a pulsing behavior.
During each cycle of velocity variation, the lift starts to fluctuate when the velocity is about to reach minimum.
Such oscillations are due to the absolute instability of the wake as the instantaneous Reynolds number drops below $Re\approx 120$.
As the velocity increases from its minimum value, the fluctuation in the lift coefficient grows drastically.
However, since the perturbation in the flow does not have time grow strong enough to trigger the transition from state I to state II, the oscillatory lift coefficient peaks before the velocity reaches the maximum value, and then damps out at high instantaneous Reynolds numbers, at which the wake stays at state I.

\begin{figure}
\centering
\includegraphics[width=0.99\textwidth]{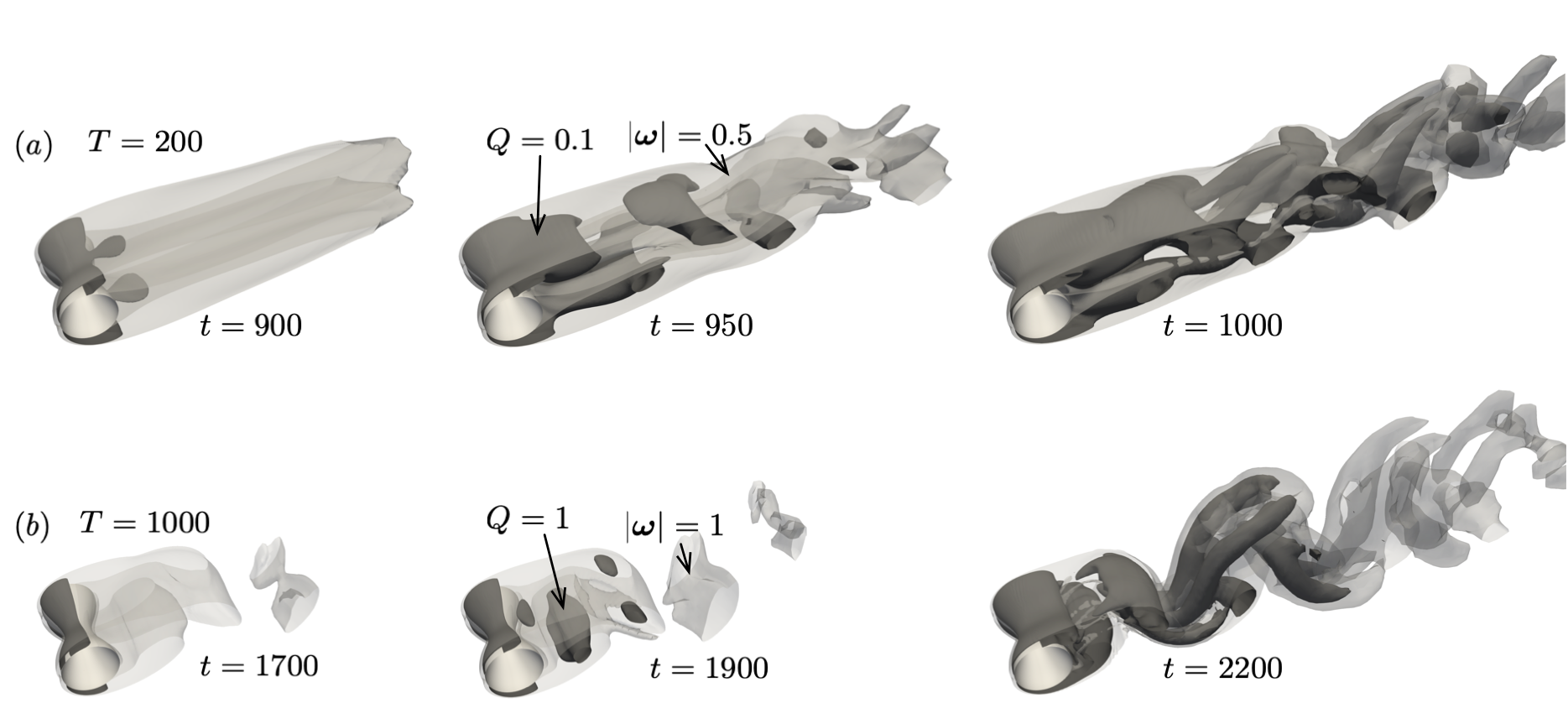}
\caption{Wake structures in the wake of the wavy cylinder under time-varying inflow velocity. $(a)$ $T=200$ and $(b)$ $T=1000$. The vortical structures are visualized by isosurfaces of $|\boldsymbol{\omega}|$ (transparent) and $Q$ (solid).}
\label{fig:gustVortical}
\end{figure}

Given long period of $T=1000$, the perturbation in the flow is allowed to grow strong enough to trigger the transition to state II, as shown in figure \ref{fig:gustCdCl}($c$).
The fluctuations in the lift coefficient are almost in phase with the inlet velocity, and are much higher than that in the $T=200$ case.
In addition, as the velocity reaches maximum, oscillations also become noticeable in the drag coefficient.
The wake vortical structures for the $T=200$ and $T=1000$ cases are shown in figure \ref{fig:gustVortical}.
In the former, the flow is featured by the typical state I wake with weak undulations.
For the latter, the wake is unsteady throughout one velocity variation cycle, and the vortical structures appear stronger with increasing instantaneous inlet velocity.
While an exhaustive investigation of the effect of the streamwise gust period $T$ on the wake behavior is out of scope for this study, the above analysis showcases the complex transition behaviors between the bistable states in an unsteady flow setting, and calls for caution in the use of wavy cylinder as a flow control device.

\section{Conclusions}
\label{sec:conclusions}

We have carried out direct numerical simulations to reveal the Reynolds number effects on the wake dynamics of a wavy circular cylinder.
Depending on the initial condition, the wake of the wavy cylinder can settle into two distinct states, i.e., one featuring low drag and minimum lift fluctuation (state I), and another featuring unsteady vortex shedding with higher drag and lift (state II).
For state I, the wake is characterized by steady tongue-like vortical structures at low $Re$.
With increasing Reynolds number, such tongue-like structures sway back and forth in the spanwise direction.
The sway motion is related to low-frequency variation in the drag force, but does not affect the lift.
For state II, a transition from single-frequency vortex shedding to dual-frequency is observed with increasing Reynolds number.
The emergence of the second peak is associated with highly three-dimensional vortical structures.
The wakes at both states transitions to turbulent flows with the emergence of small-scale vortices at higher Reynolds numbers.

In addition, we have also simulated steamwise gust flows over the wavy cylinder to shed light on the transition behaviors between the two wake states.
The time-varying inflow velocity results in a wide range of instantaneous Reynolds number spanning from the absolutely unstable flow regime to the bistable regime.
For gust period that is of the same scale with the natural shedding period of the state II wake, the unsteady shedding state revives despite the initially steady flow, and leads to quasi-periodic flow behavior.
With much higher gust period, the wake perturbations grow in the absolutely unsteady flow regime as the instantaneous Reynolds number drops below the threshold of bistability.
If the perturbations do not grow strong enough in this regime, they are damped out in the state I wake as the instantaneous $Re$ enters the bistable regime, 
However, given long enough gust period, the perturbations can grow significantly strong to trigger the transitions from state I to state II, giving rise to unsteady flow over the entire gust period.

The above analysis have revealed novel flow physics regarding the bistable states of the wavy cylinder at unexplored Reynolds numbers, and showcased the complex transition behavior between the two states in an unsteady flow setting. The insights gained from this study improve the understanding of the wake dynamics of the wavy cylinder, and could potentially guide its use in engineering applications.

\section*{Acknowledgments}

KZ is grateful for the computing resources at Amarel cluster provided through the Office of Advanced Research Computing (OARC) at Rutgers University, on which some of the simulations were carried out.
HZ acknowledges the financial support from Natural Science Foundation of China (grant no. 52101322).
DZ thanks the support from Program for Intergovernmental International S\&T Cooperation Projects of Shanghai Municipality, China (grant no. 22160710200) and the Oceanic Interdisciplinary Program of Shanghai Jiao Tong University (grant no. SL2020PT201).

\bibliography{reference}

\begin{thebibliography}{27}
\expandafter\ifx\csname natexlab\endcsname\relax\def\natexlab#1{#1}\fi
\def\au#1{#1} \def\ed#1{#1} \def\yr#1{#1}\def\at#1{#1}\def\jt#1{\textit{#1}}
  \def\bt#1{#1}\def\bvol#1{\textbf{#1}} \def\vol#1{#1} \def\pg#1{#1}
  \def\publ#1{#1}\def\arxiv#1{#1}\def\org#1{#1}\def\st#1{\textit{#1}}

\bibitem[Ahmed \& Bays-Muchmore(1992)]{ahmed1992transverse}
{\sc \au{Ahmed, A.} \& \au{Bays-Muchmore, B.}} \yr{1992}  \at{Transverse flow
  over a wavy cylinder}.  \jt{Phys. Fluids}  \bvol{4}~(9),  \pg{1959--1967}.

\bibitem[Ahmed {\em et~al.\/}(1993)Ahmed, Khan \&
  Bays-Muchmore]{ahmed1993experimental}
{\sc \au{Ahmed, A.}, \au{Khan, M.~J.} \& \au{Bays-Muchmore, B.}} \yr{1993}
  \at{Experimental investigation of a three-dimensional bluff-body wake}.
  \jt{AIAA J.}  \bvol{31}~(3),  \pg{559--563}.

\bibitem[Assi \& Bearman(2018)]{assi2018vortex}
{\sc \au{Assi, G.~R.} \& \au{Bearman, P.~W.}} \yr{2018}  \at{Vortex-induced
  vibration of a wavy elliptic cylinder}.  \jt{J. Fluids Struct.}  \bvol{80},
  \pg{1--21}.

\bibitem[Beem {\em et~al.\/}(2012)Beem, Hildner \&
  Triantafyllou]{beem2012calibration}
{\sc \au{Beem, H.}, \au{Hildner, M.} \& \au{Triantafyllou, M.}} \yr{2012}
  \at{Calibration and validation of a harbor seal whisker-inspired flow
  sensor}.  \jt{Smart Mater. Struct.}  \bvol{22}~(1),  \pg{014012}.

\bibitem[Beem \& Triantafyllou(2015)]{beem2015wake}
{\sc \au{Beem, H.~R.} \& \au{Triantafyllou, M.~S.}} \yr{2015}  \at{Wake-induced
  'slaloming' response explains exquisite sensitivity of seal whisker-like
  sensors}.  \jt{J. Fluid Mech.}  \bvol{783},  \pg{306--322}.

\bibitem[Choi {\em et~al.\/}(2008)Choi, Jeon \& Kim]{choi2008control}
{\sc \au{Choi, H.}, \au{Jeon, W.~P.} \& \au{Kim, J.}} \yr{2008}  \at{Control of
  flow over a bluff body}.  \jt{Annu. Rev. Fluid Mech.}  \bvol{40},
  \pg{113--139}.

\bibitem[Hanke {\em et~al.\/}(2010)Hanke, Witte, Miersch, Brede, Oeffner,
  Michael, Hanke, Leder \& Dehnhardt]{hanke2010harbor}
{\sc \au{Hanke, W.}, \au{Witte, M.}, \au{Miersch, L.}, \au{Brede, M.},
  \au{Oeffner, J.}, \au{Michael, M.}, \au{Hanke, F.}, \au{Leder, A.} \&
  \au{Dehnhardt, G.}} \yr{2010}  \at{Harbor seal vibrissa morphology suppresses
  vortex-induced vibrations}.  \jt{J. Exp. Biol.}  \bvol{213}~(15),
  \pg{2665--2672}.

\bibitem[Jung \& Yoon(2014)]{jung2014large}
{\sc \au{Jung, J.~H.} \& \au{Yoon, H.~S.}} \yr{2014}  \at{Large eddy simulation
  of flow over a twisted cylinder at a subcritical {R}eynolds number}.  \jt{J.
  Fluid Mech.}  \bvol{759},  \pg{579--611}.

\bibitem[Kim \& Choi(2005)]{kim2005distributed}
{\sc \au{Kim, J.} \& \au{Choi, H.}} \yr{2005}  \at{Distributed forcing of flow
  over a circular cylinder}.  \jt{Phys. Fluids}  \bvol{17}~(3),  \pg{033103}.

\bibitem[Kim \& Choi(2019)]{kim2019lock}
{\sc \au{Kim, K.-H.} \& \au{Choi, J.-I.}} \yr{2019}  \at{Lock-in regions of
  laminar flows over a streamwise oscillating circular cylinder}.  \jt{J. Fluid
  Mech.}  \bvol{858},  \pg{315--351}.

\bibitem[Kumar {\em et~al.\/}(2008)Kumar, Sohn \& Gowda]{kumar2008passive}
{\sc \au{Kumar, R.~A.}, \au{Sohn, C.-H.} \& \au{Gowda, B. H.~L.}} \yr{2008}
  \at{Passive control of vortex-induced vibrations: an overview}.  \jt{Recent
  Pat. Mech. Eng}  \bvol{1}~(1),  \pg{1--11}.

\bibitem[Lam \& Lin(2008)]{lam2008large}
{\sc \au{Lam, K.} \& \au{Lin, Y.~F.}} \yr{2008}  \at{Large eddy simulation of
  flow around wavy cylinders at a subcritical {R}eynolds number}.  \jt{Int. J.
  Heat Fluid Fl.}  \bvol{29}~(4),  \pg{1071--1088}.

\bibitem[Lam \& Lin(2009)]{lam2009effects}
{\sc \au{Lam, K.} \& \au{Lin, Y.~F.}} \yr{2009}  \at{Effects of wavelength and
  amplitude of a wavy cylinder in cross-flow at low {R}eynolds numbers}.
  \jt{J. Fluid Mech.}  \bvol{620},  \pg{195--220}.

\bibitem[Leontini {\em et~al.\/}(2011)Leontini, Jacono \&
  Thompson]{leontini2011numerical}
{\sc \au{Leontini, J.~S.}, \au{Jacono, D.~L.} \& \au{Thompson, M.~C.}}
  \yr{2011}  \at{A numerical study of an inline oscillating cylinder in a free
  stream}.  \jt{J. Fluid Mech.}  \bvol{688},  \pg{551--568}.

\bibitem[Lin {\em et~al.\/}(2016)Lin, Bai, Alam, Zhang \& Lam]{lin2016effects}
{\sc \au{Lin, Y.~F.}, \au{Bai, H.~L.}, \au{Alam, M.~M.}, \au{Zhang, W.~G.} \&
  \au{Lam, K.}} \yr{2016}  \at{Effects of large spanwise wavelength on the wake
  of a sinusoidal wavy cylinder}.  \jt{J. Fluids Struct.}  \bvol{61},
  \pg{392--409}.

\bibitem[Lyons {\em et~al.\/}(2020)Lyons, Murphy \& Franck]{lyons2020flow}
{\sc \au{Lyons, K.}, \au{Murphy, C.~T.} \& \au{Franck, J.~A.}} \yr{2020}
  \at{Flow over seal whiskers: Importance of geometric features for force and
  frequency response}.  \jt{Plos one}  \bvol{15}~(10),  \pg{e0241142}.

\bibitem[Murphy {\em et~al.\/}(2021)Murphy, Martin, Franck \&
  Lapseritis]{murphy2021other}
{\sc \au{Murphy, C.~T.}, \au{Martin, W.~N.}, \au{Franck, J.~A.} \&
  \au{Lapseritis, J.~M.}} \yr{2021}  \at{The other navy seals: Seal whiskers as
  a bio-inspired model for the reduction of vortex-induced vibrations}.  \bt{In
  {\em Recent Trends in Naval Engineering Research\/}},  \pg{pp. 139--161}.
  \publ{Springer}.

\bibitem[Owen {\em et~al.\/}(2001)Owen, Bearman \& Szewczyk]{owen2001passive}
{\sc \au{Owen, J.~C.}, \au{Bearman, P.~W.} \& \au{Szewczyk, A.~A.}} \yr{2001}
  \at{Passive control of {VIV} with drag reduction}.  \jt{J. Fluids Struct.}
  \bvol{15}~(3-4),  \pg{597--605}.

\bibitem[Schmid(2010)]{schmid2010dynamic}
{\sc \au{Schmid, P.~J.}} \yr{2010}  \at{Dynamic mode decomposition of numerical
  and experimental data}.  \jt{J. Fluid Mech.}  \bvol{656},  \pg{5--28}.

\bibitem[Schmid(2022)]{schmid2022dynamic}
{\sc \au{Schmid, P.~J.}} \yr{2022}  \at{Dynamic mode decomposition and its
  variants}.  \jt{Annu. Rev. Fluid Mech.}  \bvol{54},  \pg{225--254}.

\bibitem[Triantafyllou {\em et~al.\/}(2016)Triantafyllou, Bourguet, Dahl \&
  Modarres-Sadeghi]{triantafyllou2016vortex}
{\sc \au{Triantafyllou, M.~S.}, \au{Bourguet, R.}, \au{Dahl, J.} \&
  \au{Modarres-Sadeghi, Y.}} \yr{2016}  \at{Vortex-induced vibrations}.  \bt{In
  {\em Springer Handbook of Ocean Engineering\/}},  \pg{pp. 819--850}.
  \publ{Springer}.

\bibitem[Xu {\em et~al.\/}(2010)Xu, Chen \& Lu]{xu2010large}
{\sc \au{Xu, C.~Y.}, \au{Chen, L.~W.} \& \au{Lu, X.~Y.}} \yr{2010}
  \at{Large-eddy simulation of the compressible flow past a wavy cylinder}.
  \jt{J. Fluid Mech.}  \bvol{665},  \pg{238--273}.

\bibitem[Yoon(2005)]{yoon2005control}
{\sc \au{Yoon, J.}} \yr{2005}  \at{Control of flow over a circular cylinder
  using wake disrupter}. PhD thesis, Master's Thesis, Seoul National
  University, Korea.

\bibitem[Zhang {\em et~al.\/}(2018{\natexlab{{\em a\/}}})Zhang, Katsuchi, Zhou,
  Yamada, Bao, Han \& Zhu]{zhang2018numerical}
{\sc \au{Zhang, K.}, \au{Katsuchi, H.}, \au{Zhou, D.}, \au{Yamada, H.},
  \au{Bao, Y.}, \au{Han, Z.} \& \au{Zhu, H.}} \yr{2018{\natexlab{{\em a\/}}}}
  \at{Numerical study of flow past a transversely oscillating wavy cylinder at
  \textit{{R}e}=5000}.  \jt{Ocean Eng.}  \bvol{169},  \pg{539--550}.

\bibitem[Zhang {\em et~al.\/}(2018{\natexlab{{\em b\/}}})Zhang, Katsuchi, Zhou,
  Yamada \& Lu]{zhang2018large}
{\sc \au{Zhang, K.}, \au{Katsuchi, H.}, \au{Zhou, D.}, \au{Yamada, H.} \&
  \au{Lu, J.}} \yr{2018{\natexlab{{\em b\/}}}}  \at{Large eddy simulation of
  flow over inclined wavy cylinders}.  \jt{J. Fluids Struct.}  \bvol{80},
  \pg{179--198}.

\bibitem[Zhang {\em et~al.\/}(2017)Zhang, Katsuchi, Zhou, Yamada, Zhang \&
  Han]{zhang2017numerical}
{\sc \au{Zhang, K.}, \au{Katsuchi, H.}, \au{Zhou, D.}, \au{Yamada, H.},
  \au{Zhang, T.} \& \au{Han, Z.}} \yr{2017}  \at{Numerical simulation of vortex
  induced vibrations of a flexibly mounted wavy cylinder at subcritical
  {R}eynolds number}.  \jt{Ocean Eng.}  \bvol{133},  \pg{170--181}.

\bibitem[Zhang {\em et~al.\/}(2020)Zhang, Zhou, Katsuchi, Yamada, Han \&
  Bao]{zhang2020bistable}
{\sc \au{Zhang, K.}, \au{Zhou, D.}, \au{Katsuchi, H.}, \au{Yamada, H.},
  \au{Han, Z.} \& \au{Bao, Y.}} \yr{2020}  \at{Bistable states in the wake of a
  wavy cylinder}.  \jt{Phys. Fluids}  \bvol{32}~(7),  \pg{074112}.

\end{thebibliography}
\bibliographystyle{jfm}
\end{document}